\documentstyle[epsf,flushrt]{ptptex}

\newcommand{\simgt}{\lower.5ex\hbox{$\; \buildrel > \over \sim \;$}}
\newcommand{\simlt}{\lower.5ex\hbox{$\; \buildrel < \over \sim \;$}}
\def\wq{{w}}

\title{
Possible Method to Reconstruct the Cosmic Equation of 
State from Strong Gravitational Lensing Systems 
}

\author{
Kazuhiro {\sc Yamamoto}${}^{1,}$\footnote{E-mail : 
kazuhiro@hiroshima-u.ac.jp} 
and
Toshifumi {\sc Futamase}${}^{2,}$\footnote{E-mail : 
tof@astr.tohoku.ac.jp} 
}

\inst{
~${}^{1}$ Department of Physics, Hiroshima University, 
\\
Higashi-Hiroshima 739-8526, Japan
\\
\vspace{1mm}
~${}^{2}$ Astronomical Institute, Graduate School of Science, 
Tohoku University,\\
 Sendai 980-8578, Japan }

\recdate{January 19, 2001}

\begin{document}

\abst{%
A possible method to reconstruct the cosmic equation of state 
using strong gravitational lensing systems is proposed. 
The feasibility of the method is investigated by carrying out the 
reconstruction on the basis of a simple Monte-Carlo simulation. 
We show that the method can work and that the cosmic equation of state
$w(z)$ can be determined within errors of $\Delta w\sim \pm0.1$ -- $\pm0.2$ 
when a sufficiently large number of lensing systems ($N\sim 20$) for 
$z\simlt 1$ are precisely measured. Statistics of lensed sources in a 
wide and deep survey like the SDSS are also briefly discussed.
}

\maketitle

\section{\bf Introduction}
Because of recent progress in the capability of observational 
facilities, gravitational lensing phenomena in the high-redshift 
universe have come to play a very important role in the fields 
of cosmology and astrophysics.\cite{rf:TF} 
Gravitational lensing phenomena are useful not only to probe the
nature of dark matter in the universe but also to test cosmological 
models. Recent reports on the measurement of 
the cosmic shear field demonstrate this
usefulness.\cite{rf:vWME,rf:BRE,rf:WTK,rf:KWL,rf:MvWM}
It is also well recognized that strong lensing systems at high redshift 
are useful to constrain cosmological models, in particular, to test 
the cosmological constant in the universe.\cite{rf:FH,rf:FFK,rf:Turner}

Recent observations of cosmic microwave background anisotropies 
and distant supernovae favor a spatially flat universe whose 
expansion is currently accelerating. The cosmological 
model that includes a cosmological constant is the simplest 
model that explains these observations. Motivated by those 
observational results, variants of the cosmological constant model
have been proposed in the framework of the cold dark matter (CDM) 
cosmological model. For example, quintessence has been recently 
investigated. A model that includes quintessence has the 
attractive feature of explaining the `coincidence problem' 
-- the near coincidence of the density of 
matter and the dark energy component at present.
\cite{rf:Caldwell98,rf:Zlatev98,rf:Steinhardt99}

Such generalized cosmological constants are regarded as a dark energy 
component in addition to the dark matter component in the universe. 
The dark energy component can be characterized by the equation of
state $\wq=p_Q/\rho_Q$, where $p_Q$ is the pressure and $\rho_Q$ 
is the energy density. If the dark energy is the cosmological
constant, then $\wq=-1$. For the quintessence model, $\wq$ takes 
an arbitrary value satisfying $-1<\wq\leq0$, and is a function of 
the redshift in general.
Thus constraints on the equation of state are quite important for 
models of the dark energy component. 
Predictions of the CDM models with the dark energy component have 
been tested by comparison with various cosmological observations.
\cite{rf:Efs,rf:Per,rf:Wng,rf:ND,rf:HT}
An interesting approach to understanding the dark energy component is 
reconstruction of the cosmic equation of state or the
quintessential potential.\cite{rf:Str,rf:NC99,rf:Saini,rf:CN00}

Several authors have investigated gravitational lens systems as 
probes of the cosmic equation of state.\cite{rf:Zhu,rf:CH,rf:FY}
Recently Futamase and Yoshida proposed a possible method
to probe the dark energy component using strong gravitational
lensing systems.\cite{rf:FY} 
They have pointed out that even one lensing system might 
give a strong constraint on the time variation of the 
dark energy component. Here, one might ask whether it is possible 
to reconstruct the cosmic equation of state or the
quintessential potential using many strong lens systems, 
which may be detected using wide and deep surveys, e.g., 
the SDSS. In the present paper, motivated by the 
investigation in Ref.~\citen{rf:FY}, we investigate the feasibility 
of a scheme for reconstructing the redshift evolution of the cosmic 
equation of state.

This paper is organized as follows. In \S 2, we briefly review 
theoretical formulas of the lensing model and a background cosmological 
model. In \S 3, a reconstruction scheme is worked 
out. In \S 4, the statistics of lensed sources detected in a wide 
and deep survey like the SDSS are briefly discussed.
\S 5 is devoted to summary and conclusions. Throughout 
this paper we use units in which the velocity of light, $c$, equals $1$.

\section{\bf Lens equation}
\def\hf{{f}}

We restrict ourselves to a spatially flat FRW universe 
whose line element is written as 
\begin{equation}
ds^2=a(\eta)^2(-d\eta^2+dr^2+r^2d\Omega_{(2)}^2),
\end{equation}
where $d\Omega_{(2)}^2$ is the line element of a unit two-sphere,
$\eta$ is the conformal time and $a(\eta)$ is the scale factor 
normalized as $a=1$ at present.
We use a quintessential cosmological model consisting 
of a scalar field slowly rolling down its effect potential. 
Using the pressure $p_Q$ and the energy density $\rho_Q$ 
of the dark energy component, the effective equation of state 
of the dark energy component is written $\wq=p_Q/\rho_Q$,
which is a function of the redshift in general quintessential models. 
Assuming the equation of motion 
\begin{equation}
   d(\rho_Q a^3)+ p_Qd(a^3)=0,
\end{equation}
with the effective equation of state $\wq(z)$,
we have the solution 
\begin{equation}
   {\rho_Q(z)\over \rho_Q(z=0)}=(1+z)^3
  \exp\biggl[3\int_0^z{\wq(z')\over 1+z' }dz'\biggr]\equiv \hf(z).
\end{equation}
In the case that the equation of state is written as $\wq(z)=w_0(1+z)^\nu$,
we have
\begin{equation}
  f(z)=(1+z)^3 \exp\biggl[3w_0\biggl({(1+z)^\nu-1\over \nu}\biggr)\biggr].
\end{equation}
In the limit of a constant equation of state, $\nu=0$, 
the dark energy density evolves as $\rho_Q(z)\propto f(z)=(1+z)^{3(1+w_0)}$.
We denote the density parameters of the matter component
and the dark energy at present by $\Omega_0$ and 
$\Omega_Q(=1-\Omega_0)$, respectively.  Then, from the Friedman 
equation, we have the formula for the angular diameter
distance
\begin{eqnarray}
  D_A(z_1,z_2)={1\over H_0(1+z_2)}\int_{z_1}^{z_2}
  {dz'\over
  [\Omega_0(1+z')^3+\Omega_Q \hf(z')]^{1/2}},
\label{DA}
\end{eqnarray}
where we assumed $z_1<z_2$, and $H_0=100h~{\rm km/s/Mpc}$ is 
the Hubble constant.

Following previous investigations,~\cite{rf:FH,rf:FY} we here
consider the lens equation with a lens potential
given by an isothermal ellipsoid model  as\cite{rf:KSB}
\begin{equation}
  \Phi=4\pi\sigma_v^2{D_{LS}\over D_{S}}
  \sqrt{(1-\epsilon)\theta_1^2+(1+\epsilon)\theta_2^2},
\end{equation}
where $\sigma_v$ is the one-dimensional velocity dispersion,
$D_{LS}$ is the angular diameter distance between lens and 
source objects, and $D_{S}$ is the distance between the source 
and the observer. 
The ratio $e$ of the minor axis to the major axis is related 
to the ellipticity $\epsilon$ by $e=\sqrt{(1+\epsilon)/(1-\epsilon)}$.
Thus the lens equation gives the elliptical image of
the Einstein ring with minor axis and major axis as 
\begin{equation}
 \theta_{\pm} = \theta_E \sqrt{1 \pm \epsilon},
\label{DD}
\end{equation}
where $\theta_E=4\pi \sigma^2_v D_{LS}/ D_S$.
Denoting the redshifts of the lens and the source by $z_l$ and $z_s$,
respectively, we can write $D_{LS}=D_A(z_l,z_s)$ 
and $D_{S}=D_A(0,z_s)$.

\section{\bf Reconstruction scheme}
\def\rmf{{\rm f}}

We assume a set of  lensing systems whose observable 
quantities $z_l$ and $z_s$ and $\sigma_v,~\epsilon$ and $\theta_E$ 
are determined for each system. 
If $\sigma_v,~\epsilon$ and $\theta_E$ are measured for a lensing 
system, ${D_{S}/ D_{LS}}$ is determined from Eq.~(\ref{DD}).
Thus the measurement of $\sigma_v,~\epsilon$ and $\theta_E$ determines 
the ratio of distances ${D_{S}/ D_{LS}}$. In our reconstruction
procedure, we assume that a set of observable quantities $z_l$, $z_s$ 
and $R\equiv{D_{S}/D_{LS}}$, instead of $\sigma_v,~\epsilon$ and $\theta_E$
are determined. To be specific, we construct the set of $N$ assumed 
lensing systems in a simulation as follows. First, we choose $z_s$ 
through a homogeneous 
random process in the range of redshifts 
$z_s^{\rm min}\leq z_s\leq z_s^{\rm max}$.
Throughout this section we set $z_s^{\rm min}=0.5$ and $z_s^{\rm max}=1$.
In a similar way, we choose $z_l$ in the range 
$z_l^{\rm min}\leq z_l\leq z_l^{\rm max}$, where we set $z_l^{\rm min}=0.1$ 
and $z_l^{\rm max}= z_s^{\rm max}-0.2$ in the present paper.
Using a cosmological model, we then compute ${D_{S}/ D_{LS}}$
using the distance-formula (\ref{DA}). Observational errors 
will cause an observed value of ${D_{S}/ D_{LS}}$ to differ
from the theoretical prediction. We assume that the ellipticity 
and the angle of the Einstein ring are well determined for each 
lensing system and that the observational noise is dominated by an 
error in $\sigma_v$. Then we set 
\begin{equation}
 R={D_{S}(z_l,z_s)\over D_{LS}(z_l,z_s)}\times (1+p)^2,
\label{defR}
\end{equation}
where $p$ is a random variable that is characterized by a Gaussian
probability function with mean $0$ and the variance $\sigma^2$.
\footnote{Note that the variance $\sigma^2$ is distinct from 
$\sigma_v^2$, which represents the velocity dispersion for the lens 
model.}

We next consider reconstructing the cosmic equation of state from 
the set of $N$ lensing systems specified by $z_{li},~z_{si}$ and $R_i$,
where the subscript $i$ denotes the data of the $i$-th lensing system. 
Our method is a simple application of the reconstruction
scheme from supernovae data proposed by Chiba and Nakamura.~\cite{rf:CN00}
We adopt the following fitting formula for the angular diameter 
distance:
\begin{eqnarray}
  &&D_{A\rmf}(z_1,z_2)={1\over H_0}{1\over 1+z_2}
  \Bigl(\eta_\rmf(z_1)-\eta_\rmf(z_2)\Bigr),
\end{eqnarray}
with 
\begin{eqnarray}
  \eta_\rmf(z)=2\alpha [y^{-8}+\beta y^{-6}+\gamma y^{-4} 
  +\delta y^{-2}+\zeta ]^{-1/8}, 
\end{eqnarray}
and $y=1/\sqrt{1+z}$. 
Here the subscript ${\rmf}$ refers to
 fitting formulas. We use the constraint equation
from the Friedman equation\cite{rf:CN00}
\begin{equation}
  \alpha={[1+\beta+\gamma+\delta+\zeta]^{9/8}
  \over 1+ 3\beta/4+\gamma/2+\delta/4}.
\label{relalp}
\end{equation}
Then we determine the parameters $\beta,~\gamma,~\delta$ and $\zeta$ 
so as to minimize 
\begin{equation}
  F_A(\beta,\gamma,\delta,\zeta)=\sum_{i=1}^N \Biggl[{(R_i-
  R_{{\rm f}i}(z_{li},z_{si},\beta,\gamma,\delta,\zeta))^2\over R_i^2}\Biggr],
\label{defA}
\end{equation}
where $R_{{\rm f}i}$ is defined by
\begin{eqnarray}
   R_{\rmf i}(z_{li},z_{si},\beta,\gamma,\delta,\zeta)=
  {D_{A\rmf}(0,z_{si})\over D_{A\rmf}(z_{li},z_{si})}
  ={\eta_\rmf(0)-\eta_\rmf(z_{si})
  \over \eta_\rmf(z_{li})-\eta_\rmf(z_{si})}.
\end{eqnarray}
Thus $R_{\rmf i}$ does not depend on $\alpha$, so the relation 
(\ref{relalp}) is needed to determine $\alpha$ in our 
prescription. Finally, the cosmic equation of state is given by 
\begin{eqnarray}
  \wq_\rmf(z)
&=&
  {-4yd^2\eta_\rmf/dy^2 \over 3(d\eta_\rmf/dy)
  \bigl(\Omega_0(d\eta_\rmf/dy)^2-4\bigr)}.
\end{eqnarray}

Figure 1 demonstrates the feasibility of the reconstruction scheme
for cosmological models whose parameters are 
$\Omega_0=0.3,~\wq_0=-0.5,~\nu=0.4$ (left panels) and 
$\Omega_0=0.3,~\wq_0=-0.7,~\nu=-0.4$ (right panels).
In each panel the dashed curve represents the theoretical
curve $\wq(z)=w_0(1+z)^\nu$, while the solid curve and shaded region 
represent the mean and the $1$-sigma variance of the ensemble average of 
the reconstructed curves. The upper panels display the results for which 
the parameters $\beta,\gamma,\delta$ and $\zeta$ are determined by 
minimizing the function $F_A$. The middle panels correspond to the
case in which the parameters are determined so as to 
minimize the function $F_B$ instead of $F_A$ , where
\begin{equation}
  F_B(\beta,\gamma,\delta,\zeta)=\sum_{i=1}^N \Biggl[
  {(R_{{\rm f}i}(z_{li},z_{si},\beta,\gamma,\delta,\zeta)-
  R_i)^2\over R_{{\rm f}i}(z_{li},z_{si},\beta,\gamma,\delta,\zeta)^2}
  \Biggr].
\label{defFB}
\end{equation}
The lower panels display the results for which the 
reconstructed curve of $w_{\rm f}(z)$ is obtained by averaging the 
two reconstructed curves obtained using $F_A$ and $F_B$.
In Fig.~1 we assumed $\sigma=0.03$ and $N=20$.

The capability of the reconstruction scheme depends on the errors 
involved in measuring the lensing systems, which are incorporated 
by $p$ (or $\sigma$) in Eq.~(\ref{defR}) in our simulation. 
Figure 2 displays the variance of
reconstructed equation of state $\Delta w(z)$ at $z=0.5$ as
a function of $\sigma$ for the same cosmological models as in
Fig.~1, i.e.,  with $\Omega_0=0.3,~\wq_0=-0.5,~\nu=0.4$ (left panels) 
and $\Omega_0=0.3,~\wq_0=-0.7,~\nu=-0.4$ (right panels).
This shows that the equation of state can be reconstructed 
within errors of $\Delta w\sim \pm0.1$ -- $\pm0.2$.

In our simulation, for a set of data obtained
through a random process, the scheme fails in reconstruction 
and gives a non-realistic value of $\wq_{\rm f}(z)$. 
We removed such a case in our simulation and considered only
the case that the reconstructed equation of state satisfies
$-3/2\leq\wq_{\rm f}(z)\leq 0$ for $0\leq z\leq 1$.
'Failed' reconstructions appear more frequently as 
$\sigma$ becomes larger than $0.03$. 
For example the `failed' reconstruction appears at the rate
of $3$\%, $20$\% and $50$\% for $\sigma=0.01$, $0.03$ and 
$0.05$, respectively.
Hence our reconstruction scheme might not be 
feasible for $\sigma\simgt 0.03$ for general models other 
than the quintessential model.

\section{\bf Discussion -- Lensing statistics }

In the previous section we have assumed a set of $N~(=20)$ lensing 
systems distributed in the range $0.5\leq z_s \leq 1$. In this 
section we give some grounds for this assumption. Lensing 
statistics have been thoroughly investigated,~\cite{FFKT}\cite{rf:TOG}
and following previous investigations, we can roughly estimate
the number of lensed sources per redshift as (see the Appendix for a 
derivation)
\begin{eqnarray}
  &&{dN(z_s)\over dz_s} = 5.8\times 10^{7} (H_0 r(z_s))^5 
  {d H_0r(z_s)\over dz_s}
  {n_{\rm gal}(z_s)\over {h^3~{\rm Mpc}^{-3}} }
\nonumber
\\
&&\hspace{3cm}  \times \biggl(
  {n_{\rm halo}\over { 10^{-2}h^3~{\rm Mpc}^{-3}} }\biggr)
  \biggl({\sigma_v \over 250~{\rm km/s}}\biggr)^4 {\rm str}^{-1},
\label{dNdz}
\end{eqnarray}
where $r(z)$ is the comoving distance at redshift $z$,
$n_{\rm gal}(z_s)$ denotes the comoving number density of
galaxies, and $n_{\rm halo}$ denotes the comoving number 
density of the lensing halo, which is assumed to be constant 
throughout the universe. In deriving Eq.~(\ref{dNdz}), 
the singular isothermal sphere model is assumed as the 
lens model, and the probability that multiple images appear 
is computed.

Figure 3 displays the expected distribution of lensed sources
$dN/dz_s$ for cosmological models with 
$\Omega_0=0.3,~\wq_0=-0.5,~\nu=0.4$ (left panels) and 
$\Omega_0=0.3,~\wq_0=-0.7,~\nu=-0.4$ (right panels).
Here we adopted a luminosity function of the APM 
galaxies\cite{rf:LPEM} fitted to the Schechter function
\begin{equation}
\phi(L)dL=\phi^*\biggl({L\over L^*}\biggr)^\kappa
\exp \biggl(-{L\over L^*}\biggr)d\biggl({L\over L^*}\biggr),
\label{phiL}
\end{equation}  
with $\phi^*=1.40\times 10^{-2}h^3~{\rm Mpc}^{-3}$, $\kappa=-0.97$,
and $M^*=-19.50+5\log_{10}h$. Then the comoving number density of
galaxies at $z$ which are brighter than the limiting magnitude
$B_{\rm lim}$ is given by
\begin{equation}
n_{\rm gal}(z,<B_{\rm lim})=\int^{\infty}_{L(B_{\rm lim},z)}\phi(L)dL
=\phi^*\Gamma[\kappa+1,x(B_{\rm lim},z)],
\end{equation}
where
$ x(B_{\rm lim},z)=
  {L(B_{\rm lim},z)/L^*}
  =[{d_L(z)/1h^{-1}~{\rm Mpc}}]^2 10^{2.2-0.4B_{\rm lim}}$,
$\Gamma[\kappa,x]$ is the incomplete Gamma function,
and $d_L(z)$ is the luminosity distance.
In each panel of Fig.~3, the three solid curves correspond to 
the cases $B_{\rm lim}=21$,~$22$, and $22.5$.
\footnote{
In the present paper we neglect the magnification bias, because 
galaxies in the range of rather small redshift, $0.5\leq z_s \leq 1$, 
are considered as lensed sources. The magnification bias enhances 
the expected number of the lensed sources. However, this
effect does not alter our conclusion qualitatively.}

The expected number of lensed objects $N$ is obtained by
integrating $dN/dz_s$. For example we have 
$N \simeq 410,~2800,~7000$ for $B_{\rm lim}=21,~22,~22.5$, respectively,
for the model with $\Omega_0=0.3,~\wq_0=-0.7,~\nu=-0.4$ within a 
solid angle of $\pi$~{\rm str}. This is 
almost the same result as that for the model with 
$\Omega_0=0.3,~\wq_0=-0.5$ and $\nu=0.4$.
We thus conclude the SDSS project in progress should detect many 
lensed objects.
Of course, the above is an over-estimation 
because the formula contains the case in which the ratio of the magnitude of 
the two lensed images is infinite. 
The actual number of `clean' lensed systems 
near the Einstein ring should be significantly smaller. If $1$\% 
of the above estimation are clean lensed systems, this number 
should be sufficient to carry out the reconstruction proposed 
in this paper.

\section{\bf Conclusion}

In this paper we have proposed a possible method to reconstruct the
cosmic equation of state using strong gravitational lensing systems.
We have investigated the feasibility by working out a
reconstruction process based on a simple Monte-Carlo simulation. 
Our result shows that the method can work when a sufficiently large 
numbers of lensing 
systems ($N\sim 20$ at $z\simlt 1$) are precisely observed.
For example, if the velocity dispersion (the lens model) can be measured 
within errors smaller than $\sim3$\%, the cosmic equation of state 
can be determined within errors $\Delta w\sim \pm0.1$ -- $\pm0.2$.
The accuracy required for the measurement of the velocity dispersion
might be quite high. However, it should be noted that such accurate 
measurements have been made using by the Keck-II 10m 
telescope.\cite{Tonry,TonryII}

We have adopted a simple model for the redshift distribution function 
of lensed sources. Specifically, in the present paper we have assumed 
a homogeneous random distribution in the range $0.5\leq z_s \leq 1$. 
%
%
The feasibility of the reconstruction scheme sensitively depends 
on the redshift distribution of the lensing system. Hence, further 
investigations of distributions of lensing systems are needed. 
In general, the capability of our reconstruction scheme 
becomes worse as the redshift of lensed sources becomes higher. 
This feature originates from the fact that the fraction 
of the dark energy component becomes smaller relative 
to the matter component as the redshift becomes higher.
This makes $D_A(z_l,z_s)$ less sensitive to the cosmic 
equation of state $w(z)$ as $z_l$ and $z_s$ become larger.

Our investigation is based on some other idealistic assumptions. 
We have assumed a flat spatial geometry of the universe. 
The crucial assumption might be that the matter fraction 
in the universe $\Omega_0$ is determined precisely.
The errors in the determination of $\Omega_0$ cause additional
errors in reconstructing $\wq(z)$. 

We have assumed a singular isothermal ellipsoid as a model of our 
lensing galaxies because of its simplicity. However, it is
not known if the  singular isothermal ellipsoid model is a 
good model for lensing galaxies. In fact, the modeling 
of lensing galaxies is
not easy, because of many theoretical and observational
ambiguities such as the existence of an anisotropic velocity
dispersion. However, there has been some
progress in this direction.
Dynamical observations and models of lensing elliptical
galaxies in the local universe have been studied by Rix et~al. 
\cite{Rix} Furthermore, the relation between the observed stellar 
velocity dispersion and the velocity dispersion associated with dark
matter has been studied in the case of singular isothermal 
sphere model.\cite{Kochanek}
It is expected that such studies will provide more accurate results
in the near future, based on the large sample of galaxies  
in SDSS. Thus we hope that our method will arrow for realistic 
reconstruction of the cosmic equation of state in the near future.

\section*{\bf Acknowledgements}
One of the authors (K.Y.) thanks Y.~Kojima, J.~Ogura and K.~Tsunoda
for useful conversations on this topic. This research was supported 
by the Inamori Foundation and, in part, by a Grants-in-Aid from the 
Ministry of Education, Science, Sports and Culture of Japan (11640280).

\vspace{1cm}
\appendix

\section{\bf Lensing Statistics}

Here we briefly review the estimation of lensing statistics.
\cite{rf:CH,rf:TOG}
Introducing the cross section $\pi a_{\rm cr}^2$, the
optical depth due to the gravitational lensing can be written
\begin{equation}
  d\tau= n_{\rm halo}(z_l)(1+z_l)^3 \pi a_{\rm cr}^2 dt,
\end{equation}
where $n_{\rm halo}(z_l)$ is the comoving number density
of the halo at redshift $z_l$. Adopting a singular isothermal sphere 
as the model of the lens halo, 
we have 
\begin{equation}
d\tau = 16\pi^3 \sigma_v^4 n_{\rm halo}(z_l)
  \biggl({r(z_s)-r(z_l)\over r(z_s)}\biggr)^2
  r(z_l)^2dr,
\label{dtau}
\end{equation}
where $r(z)$ is the comoving distance and $\sigma_v$ is the
velocity dispersion. In deriving Eq.~(\ref{dtau}), 
we have used the assumption of a spatially flat universe and
the fact that $a_{\rm cr}$ is written as
\begin{equation}
  a_{\rm cr}=(1+z_l)^{-1}r(z_l)4\pi\sigma_v^2{D_{LS}\over D_S},
\label{defacri}
\end{equation}
where $a_{\rm cr}$ is the critical radius for multiple images. 
 
The probability function that a source at $z_s$ becomes
multiple images, which is defined by $P(z_s)=\int_0^{z_s} d\tau$, 
can be written as 
\begin{eqnarray}
  P(z_s) &=& 2.2\times 10^{-3} (H_0 r(z_s))^3 
  \biggl({ n_{\rm halo}\over { 10^{-2}h^3~{\rm Mpc}^{-3}} }\biggr)
  \biggl({\sigma_v \over 250~{\rm km/s}}\biggr)^4 ,
\label{Pzs}
\end{eqnarray}
where we have assumed the comoving number density is constant
$n_{\rm halo}(z)=n_{\rm halo}$. 
Then the expected number of lensed galaxies per unit solid angle 
is computed by
\begin{eqnarray}
  N= \int_0^{\infty} dz_s  n_{\rm gal}(z_s) P(z_s) r(z_s)^2 
  {dr(z_s)\over dz_s}  ,
\end{eqnarray}
which reduces to
\begin{eqnarray}
  N&=& 5.8\times 10^{7} \int_0^{\infty} dz_s (H_0 r(z_s))^5 
  {d H_0r(z_s)\over dz_s}
  {n_{\rm gal}(z_s)\over {h^3~{\rm Mpc}^{-3}} }
\nonumber
\\
&&\hspace{3cm}  \times \biggl(
  {n_{\rm halo}\over { 10^{-2}h^3~{\rm Mpc}^{-3}} }\biggr)
  \biggl({\sigma_v \over 250~{\rm km/s}}\biggr)^4 .
\label{dNdzA}
\end{eqnarray}

\vspace{1cm}

 \begin{figure}[t]
 \centerline{\epsfxsize=15cm \epsffile{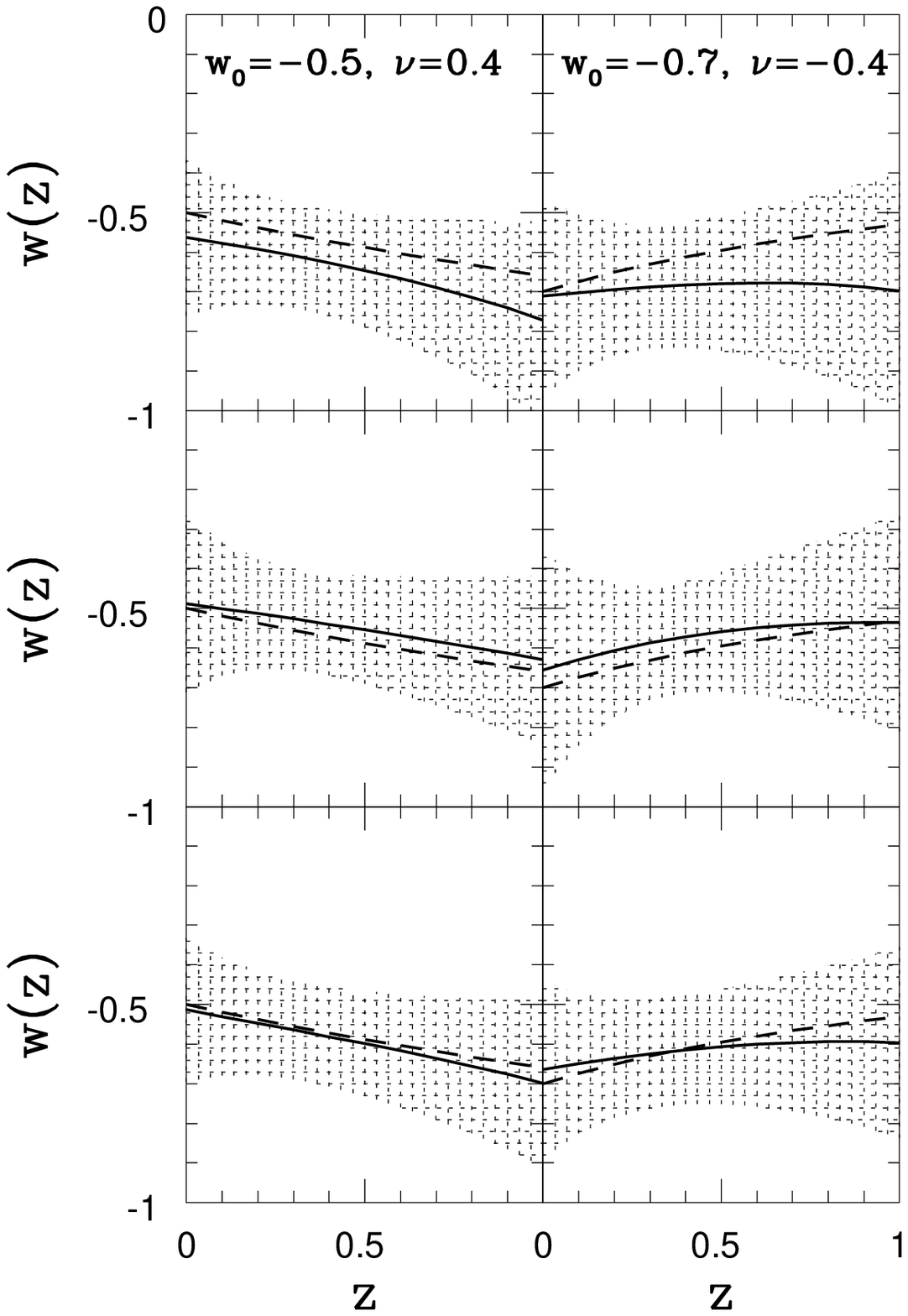}}
 \caption{Feasibility of the reconstruction scheme
for cosmological models whose parameters are
$\Omega_0=0.3,~\wq_0=-0.5,~\nu=0.4$ (left panels) and 
$\Omega_0=0.3,~\wq_0=-0.7,~\nu=-0.4$ (right panels).
Here we have assumed lensed sources ($N=20$) distributed 
in the range $0.5\leq z_s\leq 1$ and $\sigma=0.03$.
In each panel, the dashed curve is the theoretical
curve, while the solid curve and shaded region represent the mean 
and the $1$-sigma variance of the reconstructed curves. 
The upper~(middle) panels display the results for which the parameters 
$\beta,\gamma,\delta$  and $\zeta$ are determined by 
minimizing the function $F_A$~($F_B$). The lower panels 
display the result for which the reconstructed curve $w_{\rm f}(z)$ 
is obtained by averaging the reconstructed curves obtained 
from both $F_A$ and $F_B$.
}
 \end{figure}

 \begin{figure}[b]
 \centerline{\epsfxsize=15cm \epsffile{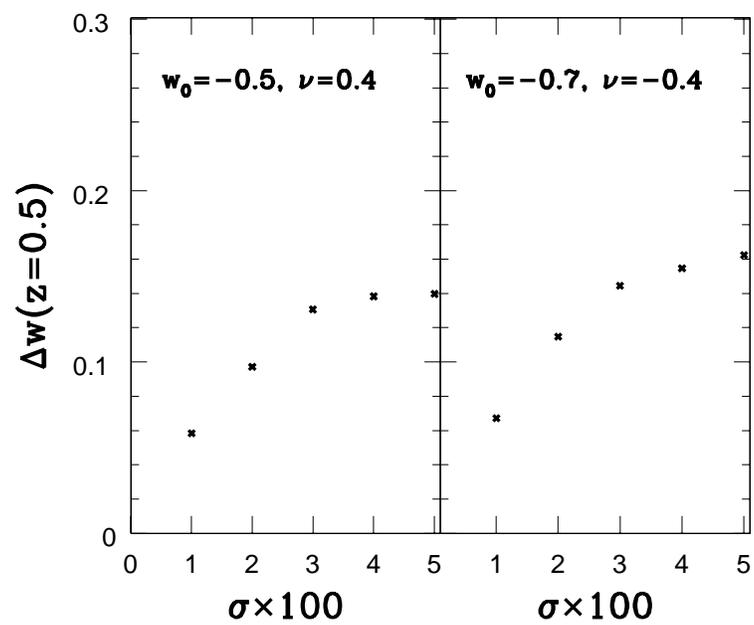}}
 \caption{
  Variance of the reconstructed cosmic equation of state $\Delta w$
  at $z=0.5$ for cosmological models whose parameters are
  $\Omega_0=0.3,~\wq_0=-0.5,~\nu=0.4$ (left panels) and 
  $\Omega_0=0.3,~\wq_0=-0.7,~\nu=-0.4$ (right panels). Here
  we have assumed that the number of lensing sources $N=20$ 
  is distributed in the range $0.5\leq z_s\leq 1$, as in Fig.~1.
}
\end{figure}

 \begin{figure}[b]
 \centerline{\epsfxsize=15cm \epsffile{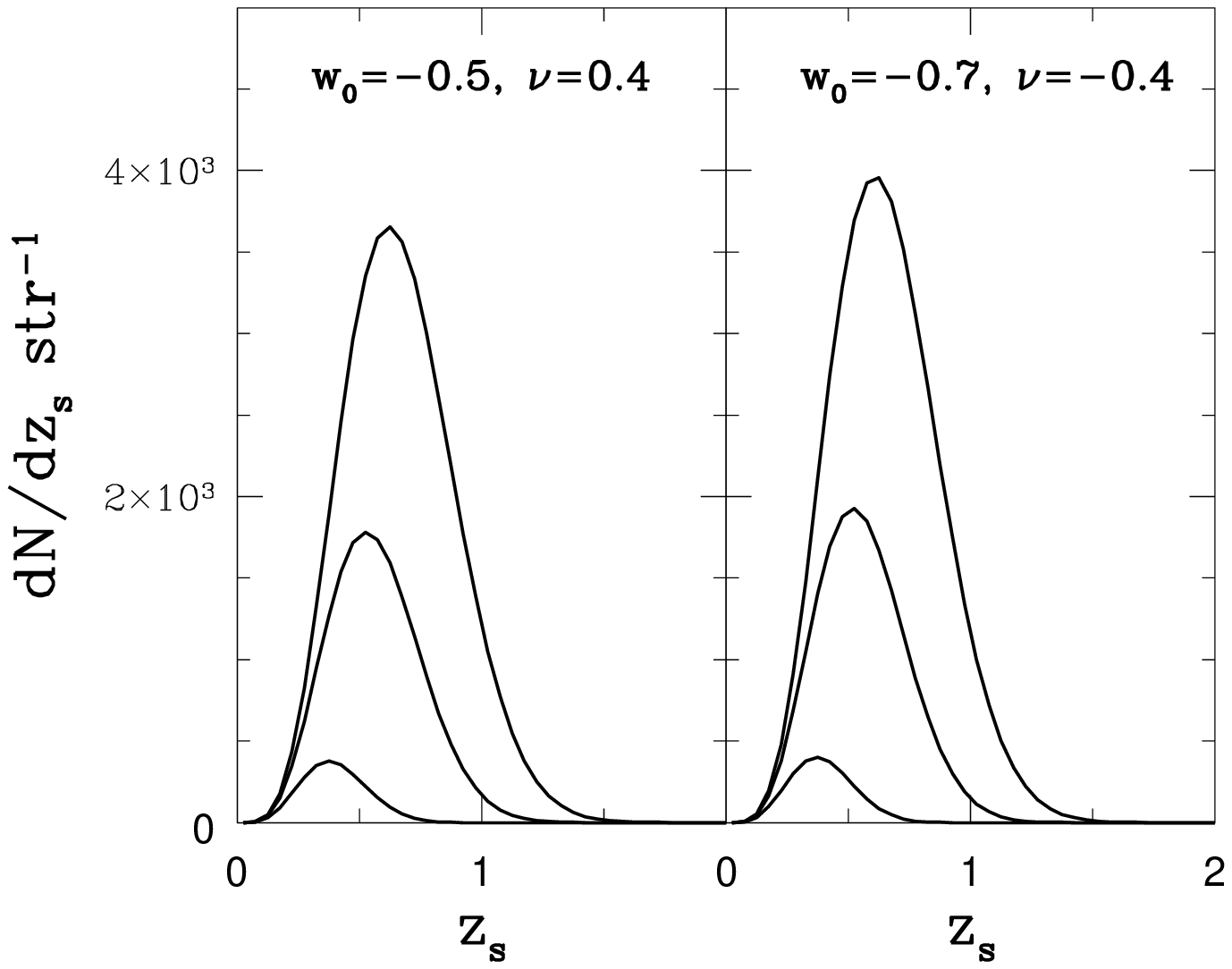}}
 \caption{
$dN/dz_s$ for cosmological models with 
$\Omega_0=0.3,~\wq_0=-0.5,~\nu=0.4$ (left panels) and 
$\Omega_0=0.3,~\wq_0=-0.7,~\nu=-0.4$ (right panels).
Here we have adopted the luminosity function for galaxies
fitted to the Schechter function. 
In each panel, the solid curves correspond to the cases 
in which the limiting magnitude is
$B_{\rm lim}=21$,~$22$ and $22.5$, from bottom to top.
}
\end{figure}


\begin{thebibliography}{99}
\bibitem{rf:TF}
  As for a review, see e.g., 
  Prog. Theor. Phys. Suppl. No.~133 (1999).
\bibitem{rf:vWME} L.~Van Waerbeke et al., Astron. Astrophys. 
{\bf 358} (2000), 30.
\bibitem{rf:BRE} D.~Bacon, A.~Refregier and R.~Ellis, 
  Mon.~Not.~R.~Astron.~Soc. {\bf 318} (2000), 625.
\bibitem{rf:WTK} D.~M.~Wittman, A.~J.~Tyson, D.~Kirkman, I.~Dell'Antonio
  and G.~Bernstein, Nature {\bf 405} (2000), 143.
\bibitem{rf:KWL} N.~Kaiser, G.~Wilson and G.~Luppino, astro-ph/0003338.
\bibitem{rf:MvWM} R.~Maoli, L.~Van Waerbeke, Y.~Mellier, P.~Schneider,
  B.~Jain, F.~Bernardeau, T.~Erben and B.~Fort, astro-ph/0011251.
\bibitem{rf:FH}
  T.~Futamase and T.~Hamana, \PTP{102,1999,1037}.
\bibitem{rf:FFK}
  M.~Fukugita, T.~Futamase and M.~Kasai, Mon.~Not.~R.~Astron.~Soc. 
 {\bf 246} (1990), L24.
\bibitem{rf:Turner}
  E.~L.~Turner,  Astrophys. J. {\bf 365} (1990), L43.
\bibitem{rf:Caldwell98}
  R.~R.~Caldwell, R.~Dave and P.~J.~Steinhardt, \PRL{80,1998,1582}.
\bibitem{rf:Zlatev98}
  I.~Zlatev, L.~Wang and P.~J.~Steinhardt, \PRL{82,1998,896}.
\bibitem{rf:Steinhardt99}
  P.~J.~Steinhardt, L.~Wang and I.~Zlatev, \PR{D59,1999,123504}.
\bibitem{rf:Efs}
  G.~Efstathiou, Mon.~Not.~R.~Astron.~Soc. {\bf 310} (1999), 842.
\bibitem{rf:Per}
  S.~Perlmutter et al. Astrophys. J. {\bf 517} (1999), 565.
\bibitem{rf:Wng}
  L.~Wang, R.~R.~Caldwell, J.~P.~Ostriker and P.~J.~Steinhardt, 
  Astrophys. J. {\bf 530} (2000), 17.
\bibitem{rf:ND}
  J.~A.~Newman and M.~Davis,  Astrophys. J. {\bf 534} (2000), L11.
\bibitem{rf:HT}
  D.~Huterer and M.~S.~Turner, astro-ph/0012510.
\bibitem{rf:Str}
 A.~A.~Starobinsky, JETP Lett. {\bf 68} (1998), 757.
\bibitem{rf:NC99}
 T.~Nakamura and T.~Chiba, Mon.~Not.~R.~Astron.~Soc. {\bf 306} (1999), 696.
\bibitem{rf:Saini}
 T.~D.~Saini, S.~Raychaudhury, V.~Sahni and A.~A.~Starobinsky,
   \PRL{85,2000,1162}.
\bibitem{rf:CN00}
 T.~Chiba and T.~Nakamura, Phys.~Rev.~{\bf D62} (2000), 121301.\\ 
 T.~Nakamura and T.~Chiba, Astrophys. J. {\bf 550} (2001), 1.
\bibitem{rf:Zhu}
  Z.~H.~Zhu, astro-ph/0010351.
\bibitem{rf:CH}
  A.~R.~Cooray and D.~Huterer, Astrophys.~J. {\bf 513} (1999), L95.
\bibitem{rf:FY}
  T.~Futamase and S.~Yoshida, gr-qc/0011083, Prog. Theor. Phys., in press.
\bibitem{rf:KSB}
  R.~Kormann, P.~Schneider and M.~Bartelmann, Astron. Astrophys. 
{\bf 284} (1994), 285.
\bibitem{FFKT}
  M.~Fukugita, T.~Futamase, M.~Kasai and E.~L.~Turner, Astrophys.~J. 
  {\bf 393} (1992), 3.
\bibitem{rf:TOG}
  E.~L.~Turner, J.~P.~Ostriker and J.~R.~Gott III, 
   Astrophys.~J. {\bf 284} (1984), 1.
\bibitem{rf:LPEM}
  J.~Loveday, B.~A.~Peterson, G.~Efsathiou and S.~J.~Maddox,
  Astrophys.~J. {\bf 390}  (1992), 338.
\bibitem{Tonry}
  J.~L.~Tonry, Astoron. J. {\bf 115} (1998), 1.
\bibitem{TonryII}
  J.~L.~Tonry and M.~Franx, Astrophys.~J. {\bf 515} (1999), 512. 
\bibitem{Rix}
  H.~W.~Rix, P.~T.~De Zeeuw, N.~Cretton, R.~P.~Van Der Marel and
  M.~Carollo, Astrophys.~J. {\bf 488} (1997), 702.
\bibitem{Kochanek}
  C.~S.~Kochanek, Astrophys.~J. {\bf 466} (1996), 638.
\end{thebibliography}
\end{document}